\documentclass[a4paper,
              ]{jacow}
\makeatletter%
	\ifboolexpr{bool{xetex}}
	 {\renewcommand{\Gin@extensions}{.pdf,%
	                    .png,.jpg,.bmp,.pict,.tif,.psd,.mac,.sga,.tga,.gif,%
	                    .eps,.ps,%
	                    }}{}
\makeatother

\ifboolexpr{bool{xetex} or bool{luatex}} 
 {}                                      
 {\usepackage[utf8]{inputenc}}           

\usepackage[USenglish]{babel}			 

\usepackage{multirow}
\usepackage{ragged2e}
\usepackage{array}
\usepackage{xcolor, soul, colortbl}
\usepackage{amssymb,amsmath,bm}
\usepackage{mathtools}
\usepackage{tabularx,setspace,booktabs,boldline,scrextend}
\usepackage{epstopdf}
\usepackage{xcolor, soul, colortbl}
\usepackage{geometry}     
\newcolumntype{C}{>{\centering\arraybackslash}X}
\usepackage{fancyhdr}
\ifboolexpr{bool{jacowbiblatex}}%
 {%
  \addbibresource{jacow-test.bib}
  \addbibresource{biblatex-examples.bib}
 }{}
\renewenvironment{thebibliography}[1]{%
   \begin{oldthebibliography}{#1}%
     \setlength{\itemsep}{-1ex}%
}%
{%
   \end{oldthebibliography}%
}

\definecolor{gray}{gray}{0.8}
\definecolor{LightCyan}{rgb}{0.95,1,1}
\definecolor{White}{rgb}{1,1,1}
\begin{document}
\newgeometry{margin = 2 cm,headheight=40pt,headsep=10pt}
\title{Beam blowup due to synchro-beta resonance with/without beam-beam effects\thanks{Work supported by JSPS KAKENHI Grant Number 17K05475. Also supported by the European Commission under Capacities 7th Framework Programme project EuCARD--2, grant agreement 312453, and under the Horizon 2020 Programme project CREMLIN, grant agreement 654166.}}

\author{K. Oide\thanks{katsunobu.oide@cern.ch}, KEK, Tsukuba, Japan \\
D. El Khechen, CERN, Geneva, Switzerland}	

\maketitle

\vspace{-2mm}
\begin{abstract}
A blowup of vertical emittance has been observed in particle tracking simulations with beam-beam and lattice misalignments\cite{ref:Dima}. It was somewhat unexpected, since estimation without lattice errors did not predict such a blowup unless a residual vertical dispersion at the interaction point (IP) is larger than a certain amount. Later such a blowup has been seen in a tracking of lattices without beam-beam effect. 

A possible explanation of the blowup is given by a Vlasov model for an equilibrium of quadratic transverse moments in the synchrotron phase space. This model predicts such a blowup as a synchro-beta resonance mainly near the first synchrotron sideband of the main $x$-$y$ coupling resonance line.
\end{abstract}

\section{Introduction}
\begin{figure}[ht!]
  \centering
  \includegraphics[width=0.45\textwidth]{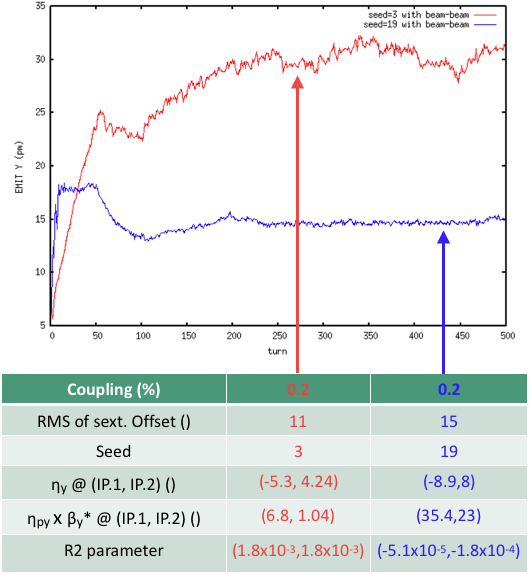}
  \caption{Blowup of vertical emittance measured at he IP by a particle tracking with beam-beam, beamstrahlung, and lattice. The arc sextupoles are vertically misaligned randomly to produce the vertical emittance of the design ratio $\varepsilon_y/\varepsilon_x=0.2\%$. Two examples for different seeds are shown, corresponding residual vertical dispersions at the IP in the table.}
  \label{fig:blowup}
\end{figure}

Beam-beam simulations with lattice, with misalignments or $x$-$y$ coupling sources such as skew quadrupoles are important to estimate the beam lifetime and luminosity evolution under more realistic situation. Such simulations have been tried for FCC-ee collider rings at t$\overline{\rm t}$ energy, 182.5~GeV. As a result, significant blowups are seen, and the magnitudes depend on the random number for the misalignments of sextupoles to generate the vertical emittance. Figure~\ref{fig:blowup} shows an example of such a blowup for two seeds of random numbers of misalignments of arc sextupoles. Note that the residual dispersion at the IP for seed 3 is smaller than the previous criteria given in Table~\ref{tab:disp}, while giving even larger blowup than another seed 19, which has larger dispersions at the IP.
\begin{table}[h!]
\begin{minipage}{8cm}
\caption{\label{tab:disp} Tolerances for residual dispersions at the IP for each energy of FCC-ee, obtained by quasi strong-weak model without lattice given by D.~Shatilov\cite{ref:fccbd}. The tolerance $\varDelta\eta_y^*$ corresponds to 5\% increase of vertical beam size $\sigma_y^*$ at the IP with beamstrahlung.}
\centering
\begin{tabular}{|l|c|c|c|c|}
\toprule
Beam energy [GeV]&45.6 & 80 & 120 & 175\\
\midrule
Design $\sigma^*_y$ [nm] & 28 & 41 & 35 & 66\\
Energy spread\footnote{with beamstrahlung} [\%]& 0.13 & 0.13 & 0.165 & 0.185\\
$\varDelta\eta_y^*$ [$\mu$m]& 1 & 5 & 4 & 6\\
\bottomrule
\end{tabular}
\end{minipage}
\end{table}

Such a blowup was somewhat unexpected, since the residual dispersion at the interaction point (IP) was not very large compared to the criteria given by beam-beam simulations without lattice. As well, simulations of beam-beam with lattice but without misalignments or skew quads, did not show such blowups\cite{ref:Zhou}.

\begin{figure*}[t!]
  \centering
  \includegraphics[width=\textwidth]{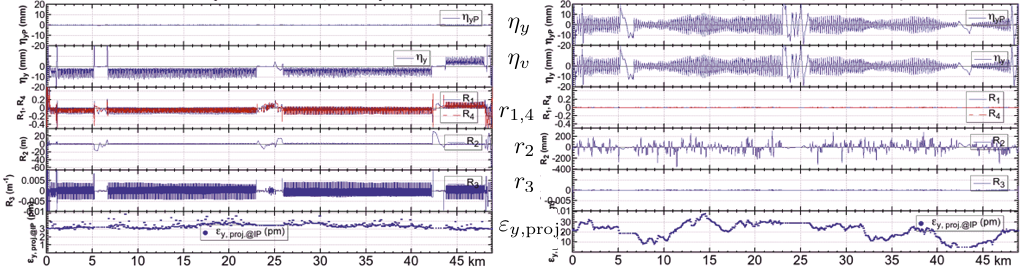}
  \caption{Comparison of beam optics between symmetric (left, $s=+1$) and antisymmetric (right, $s=-1$) skew excitations for a particular seed of random number. The columns are physical vertical dispersion $\eta_y$, dispersion in the normal mode $\eta_v$, coupling parameters $r_{1,4}$, $r_3$, $r_4$,  and the projected vertical emittance all round the half-ring. Note that the vertical emittance is generated mainly by $x$-$y$ coupling by the symmetric skew, and dispersion by antisymmetric.
  \label{fig:optics}}
\end{figure*}

\section{Method and Setup}
First let us describe the method to examine the effect in this paper:
\begin{itemize}
\item Lattice: {\tt FCCee\_t\_217\_nosol\_2.sad}, 182.5 GeV, half ring is simulated assuming a perfect period 2 periodicity. Machine parameters are listed in Ref.~\cite{ref:fccbd}.
\item The vertical emittance around the closed orbit is generated by skew quadrupole added on each sextupole in the arc. Their magnitudes on a pair of sextupoles with the $-I$ transformation between them are parametrized as
\begin{equation}
(k_1+sk_2, k_2+sk_1)\ ,
\end{equation}
where $k_1,2$ are two random numbers and $s$ is a parameter to represent the symmetry. Then $s=+1/-1$ correspond to perfect symmetric/antisymmetric excitations, and $s=0$ simply random. Examples of the  resulting optics are shown in Fig.~\ref{fig:optics} for $s=\pm1$\footnote{The definition of the coupling parameters are so defined that the uncoupled betatron coordinate is written as\begin{equation}
  \left(\begin{matrix}u\\p_u\\v\\p_v\end{matrix}\right)={\rm R}\left(\begin{matrix}x\\p_x\\y\\p_y\end{matrix}\right)=
\left(\begin{matrix}\mu&.&-r_4&r_2\\.&\mu&r_3&-r_1\\
r_1&r_2&\mu&.\\r_3&r_4&.&\mu \end{matrix}\right)\left(\begin{matrix}x\\p_x\\y\\p_y\end{matrix}\right)\ .
\end{equation}}.
\item The vertical emittance around the closed orbit is always set to $\varepsilon_y/\varepsilon_=0.2$\% unless specified otherwise.
\item Synchrotron radiation, both damping and fluctuation, is turned on in all magnets. Tapering is applied.
\item The tracking is done up to 300 half-turns with 1000 particles. The longitudinal damping is 40 turns.
\item Optionally simplified beam-beam effects and beamstrahlung are included in the Vlasov model.
\item The tracking and the Vlasov model are both done by SAD\cite{ref:SAD}.
\end{itemize}

\begin{figure}[ht!]
  \centering
  \includegraphics[width=0.45\textwidth]{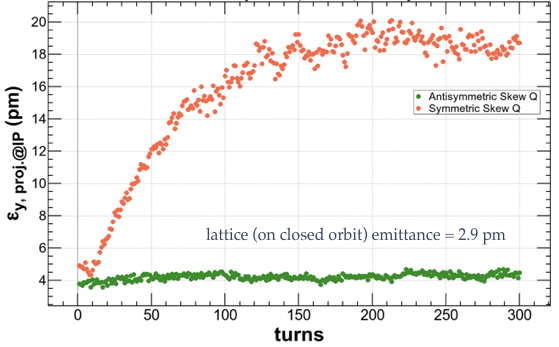}
  \caption{An example of the blowup of the vertical projected emittance at the interaction point (IP) without beam-beam for symmetric (red) and antisymmetric (green) excitations of skew quadrupoles corresponding the optcs in Fig.~\ref{fig:optics}. The blowups are quite different between them. 
  \label{fig:nobb}}
\end{figure}

Then it was found that such a blowup can occur without beam-beam effects. Figure \ref{fig:nobb} shows such an example of blowups without beam-beam. The blowup depends on the symmetry of the skew quads as well as the random number. The blowup is well explained by a Vlasov model on {\it anomalous emittance} with synchro-beta resonances introduced in Ref.~\cite{ref:ano}.

\section{Validation of the Vlasov model}
\begin{figure}[ht!]
  \centering
  \includegraphics[width=0.45\textwidth]{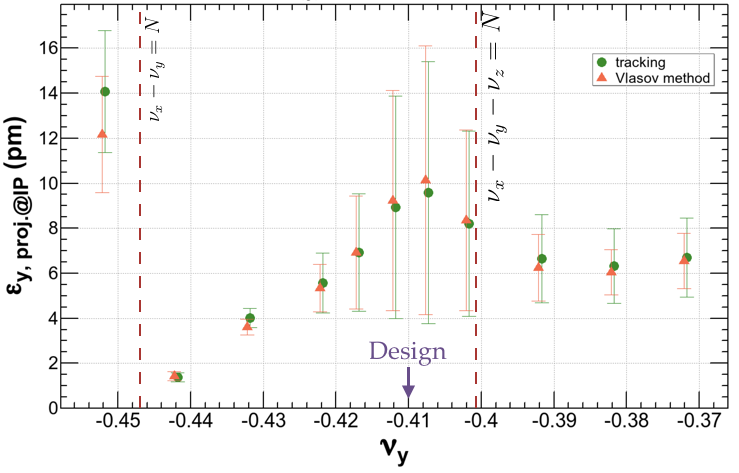}
  \caption{The blowup for different vertical tunes, by tracking (green circle) and the Vlasov model (red triangle). The error bars show the variation with 12 random numbers. Two resonances are shown by vertical dashed lines. This is done for the symmetric skew quad mode without beam-beam.
  \label{fig:nuy}}
\end{figure}

The Vlasov model in Ref.~\cite{ref:ano} obtains the equilibrium distribution of the transverse closed orbit and the quadratic beam distribution matrix in the longitudinal phase space, including the diffusion and damping by synchrotron radiation. It takes all nonlinearities of the transverse transfer matrix and the closed orbit  in momentum direction, while only linear parts in transverse planes. To see the validity of the model, let us see the dependence of the blowup against the vertical tune of the lattice. Figure~\ref{fig:nuy} shows such a tune dependence obtained by the tracking as well as the result of the Vlasov model. The agreement between them is excellent for all $\nu_y$'s in the plot,  so we will use the Vlasov mode hereafter, since it is about 1000 times faster than the tracking.

\begin{figure}[ht!]
  \centering
  \includegraphics[width=0.45\textwidth]{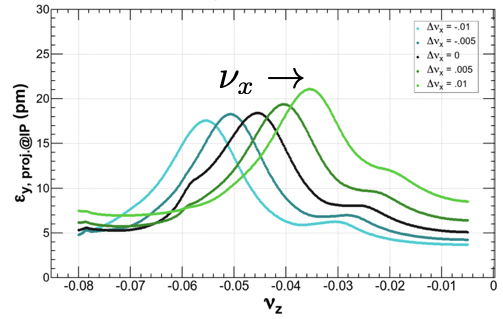}
    \includegraphics[width=0.45\textwidth]{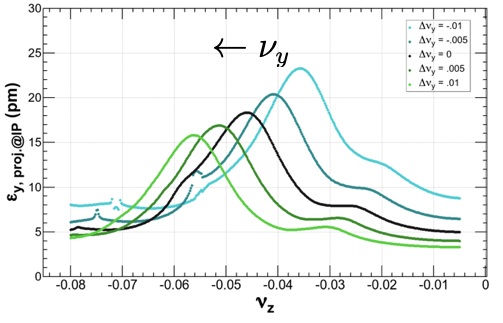}
  \caption{The blowup as a function of synchrotron tune $nu_z$ for various horizontal (upper) and vertical (lower) tunes around the design tune. The shift of the peak indicates that the most relevant resonance is $\nu_x-\nu_y-\nu_z=N$. Obtained by the Vlasov model, for symmetric skew quads, In this case, the skew quads are set to give the design vertical emittance at the design tune, and kept constant for other tunes.
  \label{fig:nuz}}
\end{figure}

\begin{figure}[ht!]
  \centering
  \includegraphics[width=0.4\textwidth]{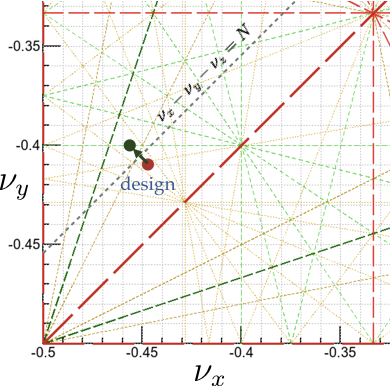}
  \caption{The tune diagram with the design transverse tune (red circle) and an alternative tune (black circle), which will be examined later. The dashed line is the first sideband of the main coupling resonance.
  \label{fig:td}}
\end{figure}

Then let us examine the dependence of the blowup by varying the synchrotron tune $\nu_z$ for several betatron tunes. Figure~\ref{fig:nuz} shows such dependences. This tune dependence clearly shows that the most relevant resonance is the first synchrotron sideband $\nu_x-\nu_y-\nu_z=N$ of the main coupling resonance, illustrated in Fig,~\ref{fig:td}.


\section{Simplified beam-beam effect in Vlasov model}
To estimate the beam-beam effect by the Vlasov model, we implemented a simplified beam-beam effects. It is a thin lens inserted at the IP. First the orbit is kicked at the IP by:
\begin{align}
\varDelta p_{x,y}=-k\frac{\partial U}{\partial(x,y)}\ ,
\end{align}
where $U$ is a potential by a Gaussian charge distribution obtained analytically. This has a nonlinear dependence on the closed orbit, which may be important when there is residual dispersion at the IP. Then the associated transfer matrix is written as:
\begin{align}
M_{\rm BB}=\left(\begin{matrix} 1 &0 &0 &0 & 0& 0\\
\displaystyle
-k\frac{\partial^2U}{\partial x^2} & 1 &\displaystyle -k\frac{\partial^2U}{\partial x\partial y} &0 & 0& 0\\
0 &0 &1 &0 & 0& 0\\
\displaystyle -k\frac{\partial^2U}{\partial x\partial y}&0&\displaystyle -k\frac{\partial^2U}{\partial y^2} & 1 &0 &0\\
0 &0 &0 &0 & 1& 0\\
0 &0 &0 &0 & 0& 1\end{matrix}\right)\ ,
\end{align}
where $k$ and the aspect ratio $\sigma_y/\sigma_x$ in $U$ are chosen to the matrix be consistent with beam-beam parameters $\xi_{x,y}$. The beamstrahlung is simplified by damping and excitation matrices at the IP:
\begin{align}
\varDelta M_{\rm D}&=\left(\begin{matrix} 0 &0 &0 &0 & 0& 0\\
0 &-d/2 &0 &0 & 0& 0\\
0 &0 &0 &0 & 0& 0\\
0 &0 &0 & -d/2& 0& 0\\
0 &0 &0 &0 & 0& 0\\
0 &0 &0 &0 & 0& -d\end{matrix}\right)\ ,\\
\varDelta\Sigma_{\rm BS}&=\left(\begin{matrix} 0 &0 &0 &0 & 0& 0\\
0 &0 &0 &0 & 0& 0\\
0 &0 &0 &0 & 0& 0\\
0 &0 &0 &0 & 0& 0\\
0 &0 &0 &0 & 0& 0\\
0 &0 &0 &0 & 0& \varDelta \sigma_\varepsilon^2\end{matrix}\right)\ ,
\end{align}
where $d$ and $\sigma_\varepsilon$ is the single-pass relative energy loss and spread due to beamstrahlung.
In this design, the numbers are $\xi_{x,y}=(0.0984,0.1414)$, $d=-7.3\times10^{-5}$, and $\sigma_\varepsilon=3.85\times10^{-4}$.

\begin{figure}[ht!]
  \centering
  \includegraphics[width=0.45\textwidth]{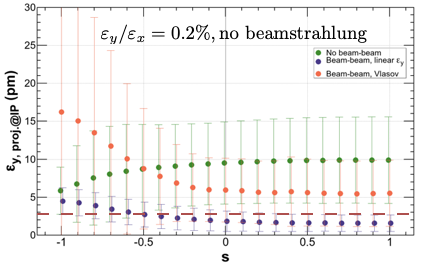}
    \includegraphics[width=0.45\textwidth]{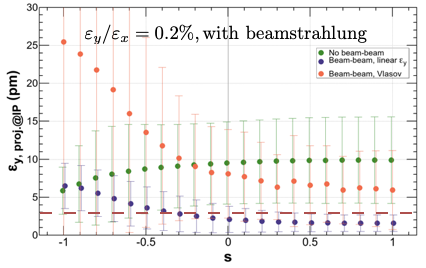}
  \caption{The blowups as a function of the symmetry parameter $s$ with/without beam-beam, without (upper) /with (lower) beamstrahlung. The error bars correspond to the variation of 12 random numbers. The design vertical emittance is shown by the dashed horizontal line. $\varepsilon_y/\varepsilon_x=0.2\%$ around the closed orbit.
  \label{fig:blowbb}}
\end{figure}

Figure~\ref{fig:blowbb} shows the results of blowup at the design tune with/without beam-beam and beamstrahlung against the symmetry parameter $s$ obtained by the Vlasov model. For the antisymmeric skew, the blowup is smaller than symmetric in the case of no beam-beam. However, with beam-beam the blowup is smaller for the symmetric ones. The beam-beam shrinks the emittance due to the large beam-beam tune shift which makes the betatron tunes off resonance, at least for the symmetric case.

 \begin{figure}[ht!]
  \centering
  \includegraphics[width=0.45\textwidth]{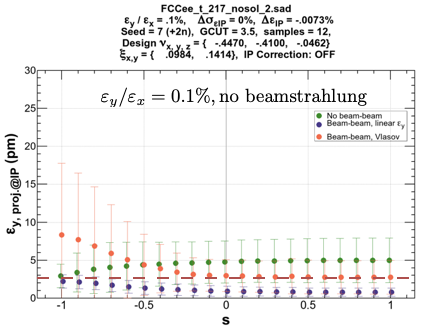}
    \includegraphics[width=0.45\textwidth]{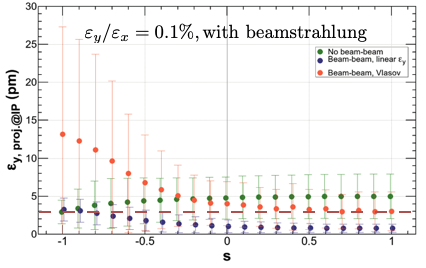}
  \caption{The blowups as a function of the symmetry parameter $s$ with/without beam-beam, without (upper) /with (lower) beamstrahlung. $\varepsilon_y/\varepsilon_x=0.1\%$ around the closed orbit.
  \label{fig:blowbb1}}
\end{figure}

The blowups with smaller vertical emittance, $\varepsilon_y/\varepsilon_x=0.1\%$, are shown in Fig.~\ref{fig:blowbb1}. The average projected emittance for $s\ge0$ barely reaches the design value. Considering the variation, The emittance without the blowup should be even smaller than 0.1\%.

\begin{figure*}[h!]
  \centering
  \includegraphics[width=0.45\textwidth]{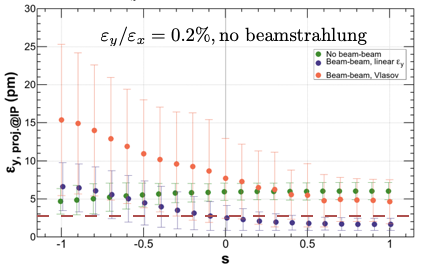}
    \includegraphics[width=0.45\textwidth]{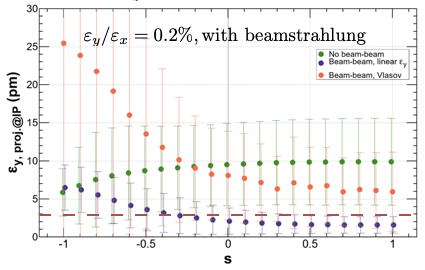}
  \caption{The blowup at the alternative tune shown in Fig.~\ref{fig:td}, with/without beam-beam, without (left) /with (right) beamstrahlung.  \label{fig:at}}
\end{figure*}

One may also shift the betatron tunes to reduce the blowup. Figure~\ref{fig:at} shows the results for the alternative tune indicated in Fig.~\ref{fig:td}. Due to other effects concerning the luminosity, the tunes are not freely chosen, and this shift, $\varDelta_{x,y}=(-0.01,0.01)$, is almost the limit.

\begin{figure*}[hb!]
  \centering
  \includegraphics[width=\textwidth]{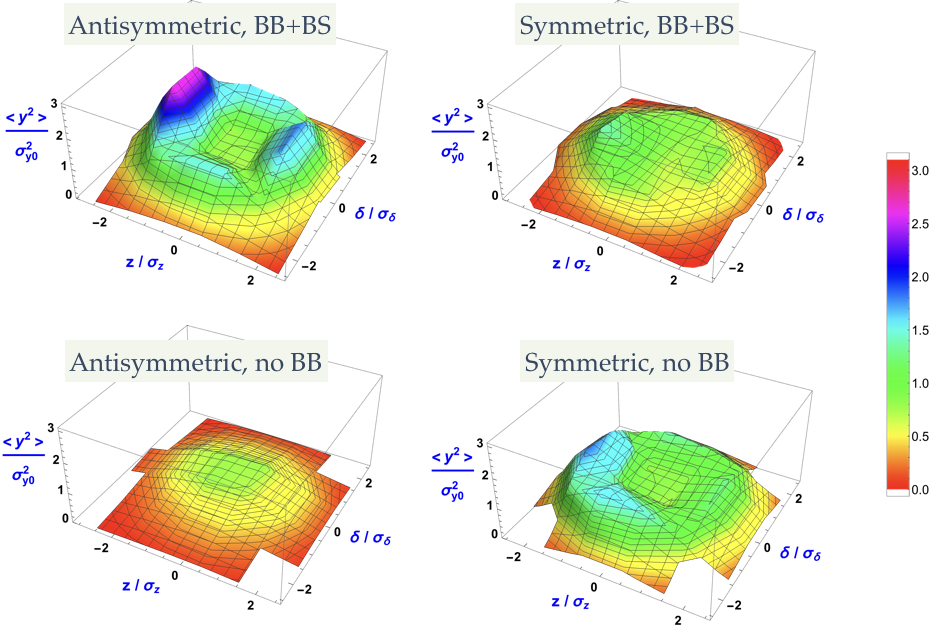}
  \caption{The  equilibrium distribution of $\langle y^2\rangle/\sigma_{y0}^2$ on the synchrotron phase space obtained by the Vlasov model. Upper left: Antisymmetric skew, with beam-beam \& beamstrahlung, upper right: symmetric skew, with beam-beam \& beamstrahlung, lower left: antisymmetric skew, without beam-beam, lower right: symmetric skew, without beam-beam.
  \label{fig:phs}}
\end{figure*}

Several distributions of $\langle y^2\rangle$ on the synchrotron phase space are shown in Fig.~\ref{fig:phs}, obtained by the Vlasov model. It is seen that large blowups occur at a certain amplitude of the synchrotron motion.

\section{summary}
The synchro-beta resonance accompanied by chromatic $x$-$y$ coupling and dispersions through the lattice of a collider causes serious beam blowup both with or without beam-beam effects and beamstrahlung. This phenomenon is well described by a Vlasov model. This will sen another criteria on the choice of the tunes and the low emittance tuning.
 
\section{acknowledgement}
The authors thank M.~Benedikt, A.~Blondel, M.~Boscolo, E.~Levichev, K.~Ohmi, D.~Shatilov, D.~Zhou, F.~Zimmermann, and the entire FCC-ee Collaboration Team for encouraging the research, useful discussions, and suggestions.

\end{document}